\documentclass[aps,prb,twocolumn,superscriptaddress,amsmath,amssymb,showpacs]{revtex4}
\usepackage{graphicx}
\usepackage{dcolumn}
\usepackage{bm}
\begin{document}

 \title{The susceptibility - Knight-shift relation in La$_{2}$NiO$_{4.17}$ and K$_2$NiF$_4$ by $^{61}$Ni NMR}
 \author{J. J. van der Klink}

 \affiliation{IPN-SB. EPFL, station 3, CH-1015 Lausanne, Switzerland}

 \author{H. B. Brom}

 \affiliation{Kamerlingh Onnes Laboratory, Leiden University, P.O.Box 9504, 2300 RA Leiden, The Netherlands}

 \date{February 25 2010, will appear in Phys. Rev. B}

\begin{abstract}
The NiO$_4$ plaquettes in ${\rm La_2NiO_{4.17}}$, a cousin of the
hole-doped high-temperature superconductor
La$_{2-x}$Sr$_x$CuO$_{4}$, have been studied by $^{61}$Ni-NMR in
14~T in a single crystal enriched in $^{61}$Ni. Doped and undoped
plaquettes are discriminated by the shift of the NMR resonance,
leading to a small line splitting, which hardly depends on
temperature or susceptibility. The smallness of the effect is
additional evidence for the location of the holes as deduced by
Sch\"{u}ssler-Langenheine {\it et al.}, Phys. Rev. Lett. {\bf 95},
156402 (2005). The increase in linewidth with decreasing
temperature shows a local field redistribution, consistent with
the formation of charge density waves or stripes. For comparison,
we studied in particular the grandmother of all planar
antiferromagnets ${\rm K_2NiF_4}$ in the paramagnetic state using
natural abundant $^{61}$Ni. The hyperfine fields in both
2-dimensional compounds appear to be remarkably small, which is
well explained by super(transferred) hyperfine interaction. In
${\rm K_2NiF_4}$ the temperature dependence of the susceptibility
and Knight shift cannot be brought onto a simple scaling curve.
This unique feature is ascribed to a different sensitivity for
correlations of these two parameters.
\end{abstract}


\maketitle

\section{Introduction}

In a seminal paper Zhang and Rice  showed the validity of a
single-band effective Hubbard hamiltonian starting from a two-band
model for the superconducting copper oxides.\cite{Zhang88}
Although the holes created by doping reside primarily on the
oxygen sites, Cu-O hybridization strongly binds this hole  to the
central Cu$^{2+}$ $3d^9$ ion of the square planar CuO$_4$
plaquette to form a singlet, nowadays called ``Zhang-Rice''
singlet. This singlet then moves through the lattice of Cu$^{2+}$
ions in a similar way as a hole in the single-band hamiltonian.

Around the same time Zaanen and Gunnarsson used a three-band
Hubbard model to describe the Cu-O perovskite plane, including the
oxygen $2p$ and Cu $3d_{x^2-y^2}$ states and argued that charge
and spins in a two dimensional electronic system with strong
correlations could form charge and spin density waves with spins
being maximal where the charges are absent.\cite{Zaanen89} This
striped structure appeared also in other theoretical
approaches,\cite{BromZaanen03} but had to wait until 1995 for a
convincing experimental verification by elastic neutron scattering
in Sr/Nd doped  ${\rm La_2CuO_4}$ and oxygen doped ${\rm
La_2NiO_4}$.\cite{Tranquada95a}

Apart from the formation of striped structures, doped cuprates and
nickelates have other similarities, but also obvious differences.
While delocalized holes in the cuprates finally lead to
superconductivity, holes in the nickelates experience strong
self-localization, and for doping levels less than one the
nickelates are insulators. Still, on the length scale of a lattice
constant the dressed holes or polarons can be seen as
non-classical objects. \cite{Pellegrin96} Like in the doped
cuprates, the holes in the nickelates are mainly located on the
in-plane oxygens, as shown for ${\rm La_{1.8}Sr_{0.2}NiO_4}$ by
resonant soft X-ray diffraction. \cite{Schussler05} Here ``undoped
Ni$^{2+}$" and ``doped Ni$^{3+}$" ions correspond to objects with
$8.2$ resp. $7.9$ $3d$ electrons, the latter accompanied by an
antiferromagnetically coupled hole in the oxygen ligand orbital of
$x^2-y^2$ symmetry.\cite{Schussler05} Since in both cases the Ni
ion is close to a $3d^8$ configuration with spin $S=1$, the doped
plaquette has a net spin of 1/2 in stead of zero spin in the
cuprates.

The present paper investigates to what extent $^{61}$Ni NMR can
give additional information about the difference between a $3d^7$
Ni$^{3+}$ ion and a doped plaquette with a (close to) $3d^8$ ion
at its center. Although the NMR technique for ${\rm
La_{2}NiO_{4.17}}$ is severely hindered by signal wipe-out due to
magnetic fluctuations below 70~K and by oxygen disordering above
250~K,\cite{AbuShiekah99,AbuShiekah01} we will argue that even in
this temperature ($T$) range two types of plaquettes can be
distinguished on the NMR time scale, and that the results are
independent evidence for the dressed $3d^8$ picture for the doped
plaquettes proposed in Ref.~\onlinecite{Schussler05}. Furthermore
we find a $T$ dependence of the linewidth that is consistent with
the formation of charge density waves or stripes, as seen in Sr
doped ${\rm La_2NiO_4}$ with a similar doping
concentration.\cite{Du00} To put our results in perspective, we
have a closer look into the paramagnetic phase of 3 and 2
dimensional (3D and 2D) antiferromagnets and in particular study
the relation between Knight shift $K$ and susceptibility $\chi$ in
the paramagnetic state of the grandmother of all square planar
antiferromagnets ${\rm K_2NiF_4}$ - one of the first studies of
this kind. Since hardly any $^{61}$Ni NMR work in non-metallic
paramagnetic systems has been published, we give some background
considerations for convenience in the Appendix.

\section{Experimental}

Several groups have investigated the location of the excess oxygen
site, possible staging and uniformity in ${\rm
La_2NiO_{4+\delta}}$ as function of $\delta$ with slightly
different results.
\cite{Rice93,Hosoya92,Tranquada94,Mehta94,Tamura96} In ${\rm
La_2NiO_{4.18}}$ the structure is (almost) tetragonal with
$2c/(a+b)=2.32$,\cite{Rice93} while the excess oxygen is located
at interstitial sites equivalent to (0.183, 0.183, 0.217).
\cite{Mehta94}

The studied single crystal (10 x 3 x 1 mm), enriched to 20 \% with
$^{61}$Ni, was grown in a mirror oven in a similar way as the
unenriched samples measured before and had similar $^{139}$La NQR
spectra. \cite{AbuShiekah99,AbuShiekah01} X-ray diffraction at
room temperature resulted in a sharp line pattern. Refinement in a
tetragonal system gives unit cell dimensions of 5.44898 \AA\ by
12.65358 \AA\, corresponding to a $c/a$ ratio of 2.322 expected
for an oxygen concentration of $1/6$ indicating a hole doping of
$1/3$.

The NMR parameters were measured with home-built NMR equipment
using standard pulse sequences by frequency sweeps at constant
field. $^{61}$Ni has $I = 3/2$, and the standard diamagnetic NMR
reference frequency based on Ni(CO)$_4$ in our nominal 14 T
magnetic field is $^{61}\omega/2\pi$ = 53.5975 MHz (a discussion
of the location of reference frequencies in general is given in
Ref. \onlinecite{VanderKlink00}, in the following referred to as
PNMRS). The high static field not only improves sensitivity, but
also separates the $^{61}$Ni quadrupole satellites from those of
$^{139}$La. \cite{AbuShiekah01} Most line positions in this paper
are given in terms of $K$,
 \begin{equation}
 K = (\omega/^{61}\omega)- 1.
 \label{e1}
 \end{equation}
\\

\begin{figure}[htb]
\begin{center}
 \includegraphics[width=8cm]{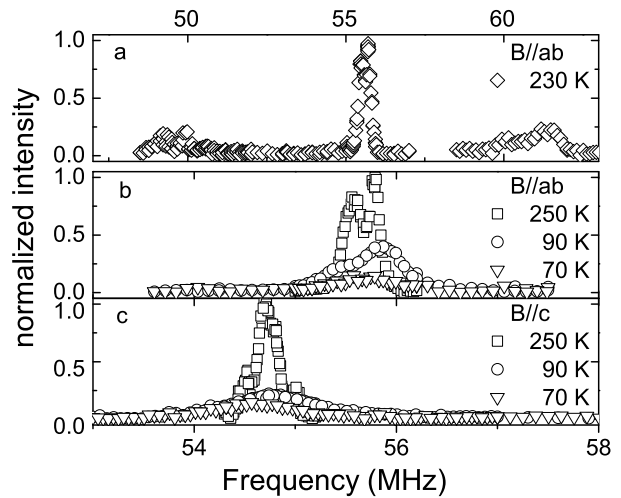}
 \end{center}
\noindent{\caption{NMR spectra for ${\rm La_2NiO_{4.17}}$ (a)
Broad frequency sweep (top axis) for $B \parallel ab$ at 230 K
showing the quadrupole splitting. (b,c) Spectra for $B
\parallel ab$ respectively $B \parallel c$ at three different
temperatures (bottom axis). If field conventions are followed, the
spectrum has to be plotted in reverse order.} \label{figfreq}}
\end{figure}

\noindent{\it La$_2$NiO$_{4+\delta}$}. Fig.~\ref{figfreq}a  gives the
results of a frequency sweep for $B \parallel ab$ between 48 and
66 MHz at 230 K revealing the Ni quadrupole splitting
 \begin{equation}
 \Delta_Q = \omega_Q(3\cos^2\theta - 1)
 \label{e2}
 \end{equation}
between the satellite transitions, with $\omega_Q/2\pi$ = 11.3
MHz. Here it has been supposed that the electric field gradient is
symmetrical around the $c$-axis, so that in Fig.~\ref{figfreq}a
$\theta = \pi/2$. In that case the central transition should be
shifted up in frequency by 0.45 MHz with respect to the center of
the two satellites. For the field parallel to the $c$-axis, the
satellites have not been observed, but should be spaced twice as
large, and the central transition should be unshifted by
quadrupolar effects. It follows that the difference in line
positions of about 1 MHz in figs.~\ref{figfreq}b and
\ref{figfreq}c can only partly be due to quadrupolar effects and a
difference in Knight shifts plays a role as well. \\

A remarkable effect is a temperature hysteresis in the
observability of these NMR signals. If the sample is cooled down
rapidly (in an hour) from room temperature to $\sim$ 200~K no
signal is visible. By cooling down slowly to 225~K, then waiting
for some hours followed by further cooling to 200~K (or  by slowly
heating to 270~K) the otherwise undetectable signals at these
temperatures can be recorded. Comparable effects have been seen in
the neutron data,\cite{Tranquada94}  and by $^{139}$La NMR in
similar crystals as studied here.\cite{AbuShiekah99} In the
$^{139}$La NMR experiment, above 230 K the $^{139}$La nuclear
relaxation is found to be primarily due to thermally activated
charge fluctuations with an activation energy of $3 \cdot 10^3$ K,
and hence the corresponding correlation rate for charge/oxygen
motion strongly depends on $T$ around ~230~K.\cite{AbuShiekah99}
By cooling faster than the correlation rate, the oxygens will be
frozen in random positions and the resulting distribution in
quadrupolar couplings will wash out the spectrum. Slow cooling
will allow the oxygens to get well-arranged. The hysteresis occurs
because the ordered dopants at low temperatures will keep their
arrangement over some $T$-range above the oxygen ordering
temperature.\cite{Tranquada94} In contrast in
La$_{5/3}$Sr$_{1/3}$NiO$_4$, which has a similar doping level but
no excess oxygen and hence only mobile holes, the transition to
charge order is of second order. \cite{Lee97,Du00,Lee01}

The intensity of the resonance lines in figs.~\ref{figfreq}b and
\ref{figfreq}c, normalized to the 250 K value, starts to decline
severely below 110~K - a phenomenon that has been observed
previously also by $^{139}$La NMR. \cite{AbuShiekah99}

\begin{figure}[htb]
 \begin{center}
 \includegraphics[width=8cm]{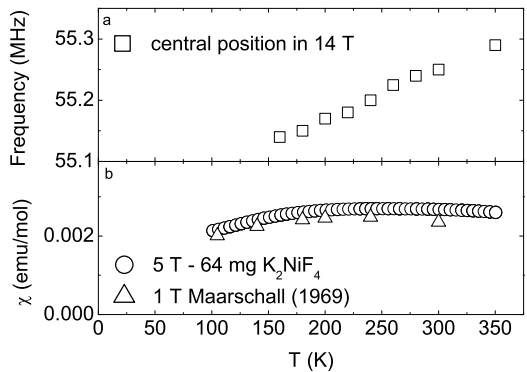}
 \end{center}
\caption{(a) $T$ dependence of the central transition of $^{61}$Ni
in a powder of ${\rm K_2NiF_4}$ in a field of 14 T (proton
frequency of 599.790 MHz) in the paramagnetic phase. (b)
susceptibility versus $T$ of the same sample (circles) and results
from Ref. \onlinecite{Maarschall69}.}
 \label{figknf1}
\end{figure}

\noindent{\it K$_2$NiF$_4$}. This compound is considered as the
prototype of a 2D Heisenberg
antiferromagnet.\cite{Maarschall69,DeJonghMiedema01} We followed
the $^{61}$Ni signal (natural abundance, powder sample) in the
paramagnetic phase as function of $T$. The central transition was
detected by Fourier-transform of half a spin-echo, and had about
25 kHz full width at half maximum (the point-charge lattice
electric field gradient is less than in the La nickelate, because
the K and F point charges are smaller than the La, resp. O). The
shift in ${\rm K_2NiF_4}$ is $T$-dependent, see Fig.
\ref{figknf1}a, and more paramagnetic (the resonance occurs at
higher frequencies or lower fields) with respect to the reference.
The clearly $T$-dependent shift does not follow $\chi$ obtained
from the magnetization measured in 5~T with a SQUID in the same
$T$ regime on the same powder as used in the NMR measurements. The
measured $\chi$ closely resembles the 1T data of Maarschall {\it
et al.},\cite{Maarschall69} see Fig. \ref{figknf1}b. The
spin-lattice relaxation time is less than a millisecond, as
expected when magnetic fluctuations are important.

\noindent{\it K$_2$NiF$_6$}. The measured $^{61}$Ni shift in an
unenriched powder of the Van Vleck paramagnet K$_2$NiF$_6$ $K =
1.08 \times 10^{-2}$ is almost $T$ independent. In this compound
Ni$^{4+}$ has a cubic environment and hence no static quadrupole
interaction; the spin-lattice relaxation times are long (of the
order of 2 seconds at 200 K), as expected when no fluctuating-spin
interactions are present.

\section{Analysis}

The NMR shift of the nuclei of paramagnetic ions in dense
paramagnets has been studied much less than that in
Pauli-paramagnetic metals (for a review of the latter, see PNMRS).
An important method of analysis of metal-NMR is the
Clogston-Jaccarino plot that correlates the spin contributions to
the susceptibility and Knight shift ($\chi_S$ resp. $K_S$) with
$T$ as implicit parameter and gives the nuclear hyperfine
field.\cite{Clogston61} The relation follows from the nuclear spin
hamiltonian (given in SI units and simplified for isotropic
interactions),
 \begin{eqnarray}
 {\mathcal{H}}_n & = & - \hbar\gamma_{\rm ref}(1 + \delta) {\vec{B}} \cdot {\vec{I}} + \hbar A {\vec{I}} \cdot <{\vec{S}}>\\ \nonumber
                 & = & - \hbar\gamma_{\rm ref}[1 + \delta + \chi_S \frac{A}{g \gamma_{ref}} \frac{\Omega}{\mu_B\mu_0}]{\vec{B}} \cdot
                 {\vec{I}}
 \label{e3}
 \end{eqnarray}
with $\Omega$ the volume of the simple unit cell, and $\delta$ the
chemical shift. The last term in the parentheses denotes the
Knight shift $K_S$. Traditionally susceptibilities are given in
the cgs unit of emu/mole, leading to
 \begin{equation}
 K_S = (\chi_{\rm mol} / N_A \mu_B) (A/\gamma_{\rm ref} g)
 \label{e4}
 \end{equation}
with $\chi_{\rm mol}$ the molar susceptibility, $N_A$ Avogadro's
number, $\mu_B$ the Bohr magneton (in cgs units), and  $g$ the
Land\'{e} $g$-factor. The combination $A/(\gamma_{\rm ref} g)$ is
referred to as the hyperfine field per Bohrmagneton, and
$A/\gamma_{\rm ref}$ as the hyperfine field per unit electronic
spin - both in Oe/Gauss. Usually $A$ is independent of $T$ so that
the Clogston-Jaccarino plot of $K_S$ versus $\chi_S$ yields a
straight line, where the slope is a direct measure of the
hyperfine field.

The $K_S$ - $\chi$ relation for the planar Cu(2) sites in ${\rm
YBa_2Cu_3O_x}$ has been analyzed by this method,
\cite{Takigawa91b,Monien91,Shimizu93} where in
Ref.~\onlinecite{Shimizu93} the relevant parameters are evaluated
in the context of a crystal-field model (see also the Appendix).
Again $\chi$ and the shift can be split into a $T$-independent
orbital and a $T$-dependent spin part. Due to the almost
tetragonal symmetry of the unit cell, the hyperfine interaction in
Eq.~\ref{e3} is anisotropic and often an extra term
$\sum_{j=1}^{4} B{\vec{S_j}}$ is added to take the transferred
interaction into account.\cite{Takigawa91b,Monien91} In underdoped
${\rm YBa_2Cu_3O_{6.63}}$ $K$ in the $c$ direction does not vary
with $T$. It means that the on-site ($T$-dependent) spin part and
supertransferred hyperfine fields cancel each other:
$A_{\parallel}+4B \approx 0$. \cite{noteTl} \\

Quite some NMR properties of paramagnetic ions in dilute
diamagnetic hosts have been derived from ENDOR (and ESR)
experiments. Two kind of results are especially relevant: often a
so-called pseudo-nuclear Zeeman effect is found, which is the
equivalent of the $T$-independent chemical shift in diamagnetic
molecules or of the Van Vleck contribution to the Knight shift in
metals;\cite{Geschwind67} furthermore supertransferred hyperfine
interactions are found on diamagnetic ions of the matrix (like
Al$^{3+}$) that are separated from the dilute paramagnetic ion
(like Fe$^{3+}$) by a ligand (like O).\cite{Taylor73} A detailed
ENDOR experiment on Ni$^{2+}$ in (trigonal) Al$_2$O$_3$  yields a
$T$-independent shift of 0.028 and a net $^{61}$Ni hyperfine field
of $\sim$ - 9 T per unit spin.\cite{Locher63}   For Co$^{2+}$ in
MgO there is a huge $T$-independent shift of $0.39 \pm
0.01$,\cite{Fry62} and a hyperfine field per unit spin of 29 T. In
the Appendix, which gives some details on the relation between NMR
and ENDOR/ESR parameters, we argue that in MgO a hypothetical NMR
experiment would see clearly distinct signals for $^{61}$Ni$^{2+}$
and $^{61}$Ni$^{3+}$.

\subsection{Metal-ion NMR in the paramagnetic phase of
antiferromagnets}

We are interested in antiferromagnetic systems for which both
ingredients for the Clogston-Jaccarino plot are available from
experiments. From the literature, these data can be found for the
3D magnets MnO (cubic) and KCoF$_3$ (perovskite); here we will add
new data on the 2D magnet K$_2$NiF$_4$.\\

\begin{figure}[htb]
 \begin{center}
 \includegraphics[width=8cm]{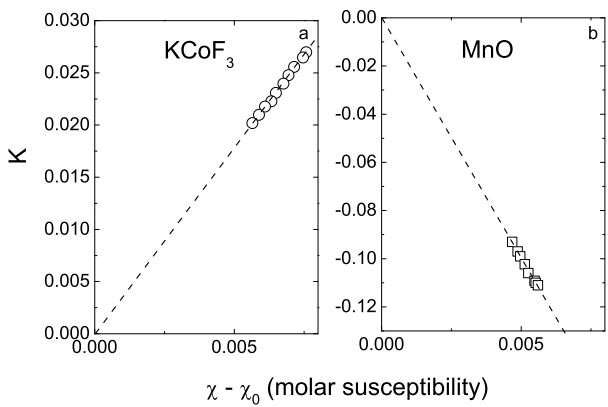}
 \end{center}
\caption{Clogston-Jaccarino plots for KCoF$_3$ based on the data
quoted in the text (a), and MnO based on data published in
Ref.~\onlinecite{Jones66} (b).}
 \label{figCJ1}
\end{figure}

\noindent{\it KCoF$_3$}. The Clogston-Jaccarino plot for KCoF$_3$
($T_N$ = 114 K) given in Fig.~\ref{figCJ1}a has not been published
before, but the $\chi$ data are available from Ref.
\onlinecite{Hirakawa64} and the NMR data from Ref.
\onlinecite{Shulman59}. We have plotted the shifts with respect to
a reference $\gamma /2\pi = 10.03$ MHz/T expected from suitably
corrected values found in ionic solutions of Co$^{3+}$ (see
PNMRS); from the $\chi$ data the $T$-independent background
$\chi_0$ has been subtracted. The rather amazing result is that
the $T$-independent shift of 0.39 from the ENDOR data is absent,
\cite{Fry62,Shulman59} although it has also been seen in the
ordered low-$T$ phase.\cite{Jaccarino59,Moriya59} A single
datapoint for paramagnetic CoO likewise fails to show this large
shift.\cite{Shulman59} The slope in Fig.~\ref{figCJ1} gives a
hyperfine field per Bohrmagneton of 20 T. Although the analysis of
the $^{59}$Co hyperfine field is very complex, \cite{Fry62} it
does not suggest that Co-Co transferred hyperfine
fields are important.\\

\noindent{\it MnO}. Also in the paramagnetic phase of MnO ($T_N$ =
117 K) transferred hyperfine fields have no noticeable influence,
see Fig.~\ref{figCJ1}b based on $^{55}$Mn NMR (after Ref.
\onlinecite{Jones66}). The extrapolation shows no $T$-independent
shifts, which is indeed expected for Mn$^{2+}$ on theoretical
grounds, see Abragam and Bleaney (referred to as
AB70).\cite{AbragamBleaney70} The slope yields a hyperfine field
per Bohrmagneton of -~11.5 T, or, using $g \sim 2$, of -~23 T per
unit spin, which goes very well with data reviewed in AB70 for
Mn$^{2+}$ diluted in simple cubic oxides, but does not support the
theoretical expectation that in MnO the hyperfine should increase
(in absolute value) by approximately 4.2 T with respect to the
value in dilute systems.\cite{Huang67} For our later discussion it
is important to remember that this
theoretically expected change is anyway relatively small.\\

\noindent{\it K$_2$NiF$_4$}. This compound ($T_N = 100.5~$K) is
the paradigm of an $S$ = 1 Heisenberg square-planar
antiferromagnet and the remarkable behavior of $\chi$ in Fig.
\ref{figknf1}b in the paramagnetic phase is theoretically well
understood.\cite{DeJonghMiedema01} It is related to the build-up
of correlations $\langle S_{z,i} S_{z,j} \rangle$ between
neighboring $(i,j)$ Ni$^{2+}$ spins. Experimentally, the existence
of such correlations has been derived by Maarschall {\it et al.}
from $T_2$ (actually linewidth) measurements on
$^{19}$F.\cite{Maarschall69} It is immediately clear from Fig.
\ref{figknf1} that no straight-line correlation  exists between
shift and susceptibility: in the experimentally accessible $T$
range the hyperfine field and $\chi$ have different
$T$-dependencies. As discussed in the Appendix we think this to be
due to a different influence of \emph{static} effects of the
correlations $\langle S_{z,i} S_{z,j} \rangle$ on the transferred
hyperfine field and on $\chi$, not seen in the experimentally
accessible $T$ regions for cuprates. This interesting effect has
no consequences for the estimate of the hyperfine field, which is
also needed for the analysis of La$_2$NiO$_4$

\begin{figure}[htb]
 \begin{center}
 \includegraphics[width=8cm]{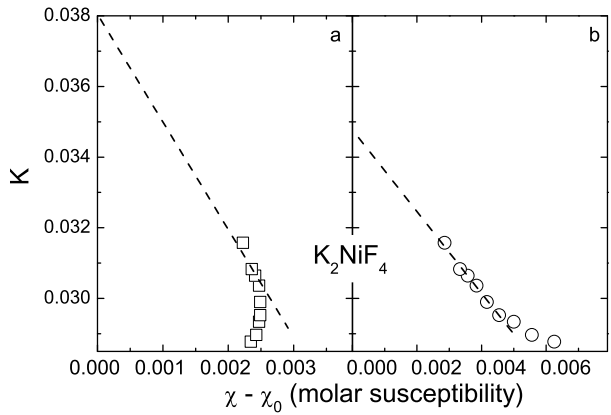}
 \end{center}
\caption{Clogston-Jaccarino plots for ${\rm K_2NiF_4}$. (a) $K$ is
plotted versus the experimental susceptibility.  (b) $K$ versus
$\chi$ of a free Ni$^{2+}$ ion with $g=2.28$. For a discussion of
the dashed lines, see text.}
 \label{figCJ2}
\end{figure}

We want to estimate the hyperfine field in the regime where $\chi$
is Curie-Weiss like. From the theoretical fits in
Ref.~\onlinecite{DeJonghMiedema01}, such behavior is expected to
occur at higher temperatures ($T > 500$~K) than we could attain in
Fig.~\ref{figknf1}. For $T > 500$~K the Clogston-Jaccarino plot
should become a straight line, and for $T^{-1} \rightarrow 0$ we
should find the $T$-independent shift. Two attempts at such a fit
are shown in Fig. \ref{figCJ2}. The $\chi$ axis in
Fig.~\ref{figCJ2}a uses the experimental data from
Fig.~\ref{figknf1}b ($\chi_0=0$), and in Fig.~\ref{figCJ2}b the
theoretical value for the Curie susceptibility of independent
spins  (again $\chi_0 = 0$) with $S$ = 1 and $g$ = 2.28  (see also
the Appendix) as found for dilute Ni$^{2+}$ in the perovskite
KMgF$_3$. \cite{Walsh63} The extrapolated values for the line
position are, somewhat arbitrarily, taken as 55.635 MHz in
Fig.~\ref{figknf1}a and 55.474 MHz in Fig.~\ref{figknf1}b
corresponding to $T$-independent contributions to the Knight shift
of $3.8 \times 10^{-2}$ and $3.5 \times 10^{-2}$, roughly three
times the value in the Van-Vleck compound ${\rm K_2NiF_6}$. The
first observation is that the slopes of both straight lines
indicate a negative hyperfine field associated with the Curie
contribution; the second that the absolute value of this field
(1.5 T in a, 0.7 T in b, expressed per Bohrmagneton) is much
smaller than the 4 T found in the ENDOR experiment.
\cite{Locher63,Geschwind67} A hyperfine field of 4 T would yield a
much steeper slope, and extrapolate to unlikely values for the
$T$-independent shift. The relatively small slopes indicate, for
the first time as far as we are aware, that in the paramagnetic
phase of a simple 2D antiferromagnet the sum of the on-site
hyperfine field and the transferred hyperfine fields can be very
small indeed. This cancellation is similar to what is found in the
cuprates, but very different from the available results for the
3D antiferromagnets in Fig.~\ref{figCJ1}.\\

\subsection{Doped La$_2$NiO$_4$}

\begin{figure}[htb]
\begin{center}
 \includegraphics[width=8cm]{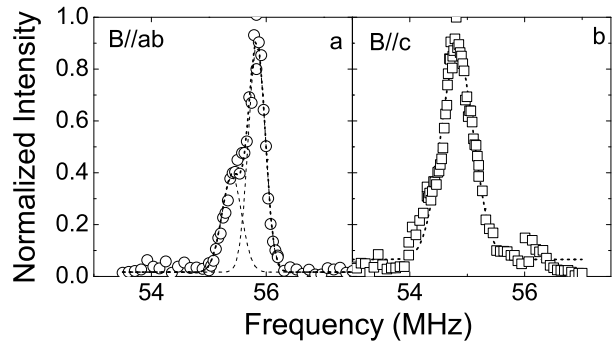}
\end{center}
\caption{Decomposition of the $^{61}$Ni line in ${\rm
La_2NiO_{4.17}}$ for $B \parallel ab$ (a) and $B
\parallel c$ (b) at 130 K. Dashed lines are Gaussian fits (see text); in (a) the two lines have an intensity ratio of 1:2.} \label{figw}
\end{figure}

The $^{61}$Ni NMR spectrum for the field ${\vec B} \parallel ab$,
see Fig. \ref{figfreq}b, has a double-peaked structure at 250~K,
that becomes a shoulder at 90~K. The line shape at 130~K is shown
in Fig. \ref{figw}a, together with a fitted decomposition into two
Gaussian lines. Apart from the shiny crystal surfaces and the
sharpness of the X-ray diffraction pattern (not shown), the
structure has to be intrinsic and not due to different crystal
domains or accidental inhomogeneities in the static distribution
of dopant oxygens for the following reasons: {\it i.} Unenriched
${\rm La_2NiO_{4.17}}$ crystals measured previously show a similar
splitting below 250~K for the $^{139}$La NMR transition between
the $-3/2$ and $-1/2$ levels and not in the $-1/2$ - $1/2$
transition, which is a proof of its quadrupolar
origin.\cite{AbuShiekah99} {\it ii.} The line shapes are very
reproducible from one $T$ run to another. Since each run starts at
high temperatures, where the oxygens diffuse rather freely, it is
unlikely that different runs end up with similar inhomogeneities.
In view of these arguments, we assign the two-peaked structure to
doped and undoped plaquettes. The intensity ratio $0.35 \pm 0.05$
to $0.65 \pm 0.05$ in Fig. \ref{figw}a is indeed as expected in
this scenario.

The $^{61}$Ni NMR data in Fig. \ref{figknf1}a are the only ones
available for what is undoubtedly Ni$^{2+}$, and from the ESR data
for Ni$^{3+}$ discussed in the Appendix the NMR shifts for
$^{61}$Ni$^{3+}$ and $^{61}$Ni$^{2+}$ are expected to differ
markedly. Hence the small line splitting in Fig. \ref{figw}a is
not due to a difference in hyperfine interaction, but has to be
ascribed to a difference in electric field gradients felt by the
Ni ion, giving additional evidence for the location of the holes
on the oxygens as deduced by Sch\"{u}ssler-Langenheine {\it et
al.}\cite{Schussler05} This electric scenario is supported by the
splitting of the $^{139}$La $m=3/2$ resonance below the same
temperature of 250~K (mentioned above), which was proven to be a
quadrupolar effect.\cite{AbuShiekah99} Also the broadening of the
spectra with decreasing $T$ has to be explained in terms of a
redistribution of charges on the surrounding oxygens (see below).
In most cuprates holes in the CuO$_2$-plane have a much higher
mobility and such a decomposition can not be made.

For ${\vec B} \parallel c$, see the discussion of Eq.~\ref{e2},
both quadrupolar and magnetic (Knight) effects can be different
from the ${\vec B} \parallel ab$ case. The line shape in Fig.
\ref{figw}b shows only the slightest hint of a low-frequency
shoulder, and for this orientation we simply fit the data to a
single Gaussian. The clearly much larger width in Fig. \ref{figw}b
is nevertheless compatible with the idea that we still have two
lines, but the accuracy of a further resolution is too small to be
meaningful.

\begin{figure}[htb]
\begin{center}
 \includegraphics[width=8cm]{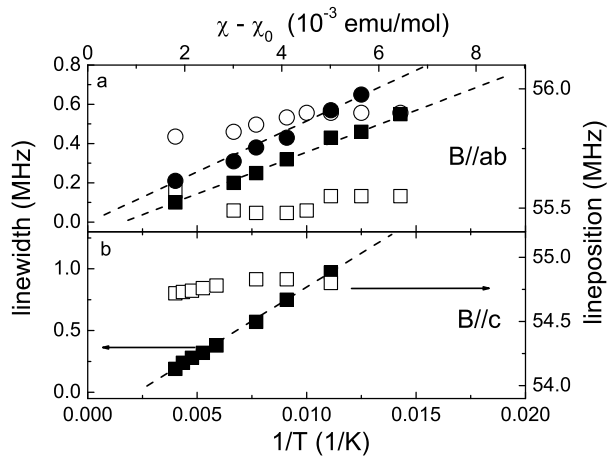}
\end{center}
\caption{Linewidth (full symbols - left axis) and position (open
symbols - right axis) of the $^{61}$Ni resonance in
La$_2$NiO$_{4.17}$ for $B\parallel ab$ (a) and $B \parallel c$ (b)
versus $1/T$ (bottom axis) and $\chi - \chi_0$ (top axis). The
reference value for the $^{61}$Ni lineposition in the applied
field is 53.60 MHz. Dashed lines are fits discussed in the text.
\label{figwT}}
\end{figure}

Clogston-Jaccarino plots of both the line positions and widths
against $\chi$ are shown in Fig. \ref{figwT}. The susceptibility
on samples with well-known oxygen stoichiometries has been
published by Odier {\it et al.},\cite{Odier99} and for samples
similar to ours by Bernal {\it et al.} and Abu-Shiekah {\it et
al.} \cite{Bernal97,AbuShiekah99} For ${\rm La_2NiO_{4.17}}$ the
$\chi$ data are in good agreement with each other, and can be
described by a Curie-Weiss law with an almost $T$-independent
background $\chi_0$.

The central positions at 54.75 MHz for $B \parallel c$ and 55.55 /
55.90 MHz for $B \perp c$ at 250~K are hardly $T$ or $\chi$
dependent. The data thus imply that the Ni-nuclei experience an
almost complete cancellation of the direct hyperfine field by the
transferred contribution, reminiscent to ${\rm K_2NiF_4}$. The
shifts with respect to the reference value of 53.60 MHz can be due
to second order quadrupolar effects (see above) and/or Van Vleck
shifts.

Below 250~K the $^{61}$Ni line grows in width with decreasing $T$,
but its intensity remains Curie like down to 110~K, where the
signal starts to get wiped-out due to magnetic
fluctuations.\cite{AbuShiekah99,Teitelbaum00,Hunt01,AbuShiekah01}
In general the line broadening can be due to a (re)distribution of
local static magnetic fields or electric field gradients, but here
$T$ is too high for a static magnetic scenario. It implies that
the Ni ions experience a increasing spread in the static (on the
NMR time scale) electrical field gradients, as expected when
charge density waves or striped structures develop. The charge
modulation affects not only the Ni-ions of the doped but also of
the undoped plaquettes. Also in unenriched ${\rm La_2NiO_{4.17}}$
crystals using $^{139}$La NMR quadrupolar effects are seen to
dominate above 140~K.\cite{AbuShiekah99} In
La$_{5/3}$Sr$_{1/3}$NiO$_4$ below the charge order temperature of
240~K a redistribution of holes has been observed
too.\cite{Lee97,Du00,Lee01}

\section{Conclusions}
In the literature on $^{63}$Cu NMR in the $2D$ cuprates it has
been accepted as an experimental fact that direct and
supertransferred hyperfine fields can cancel each other. We have
pointed out, that for such a cancellation in $3D$ systems the
available data show no indication. On the other hand, the
extrapolation of our data for ${\rm K_2NiF_4}$ in
Fig.\ref{figCJ2}, and those in Fig.\ref{figwT} for ${\rm
La_2NiO_{4.17}}$, both $2D$ antiferromagnets, suggest that this
cancellation might be a more frequent phenomenon for $2D$
systems.\\

In paramagnetic ${\rm K_2NiF_4}$ not too far above $T_N$, the
Clogston-Jaccarino relation between susceptibility and Knight
shift is not linear. This unique feature is explained by a
difference in sensitivity for correlations of these two
parameters, which probe different facets of the electronic spin. \\

Regarding the nickelates, the $^{61}$Ni NMR line position is
different for $B \parallel c$ and $B \perp c$, and doped and
undoped NiO$_4$ plaquettes can be discriminated by their line
shift found by decomposition of the resonance line for $B \perp
c$. The analysis shows that the doped holes are located on the
neighboring oxygens in agreement with the resonant soft X-ray
diffraction of Sch\"{u}ssler-Langeheine {\it et
al.}\cite{Schussler05}. From the growing line width with
decreasing temperature and the results of a previously made
$^{139}$La NMR study,\cite{AbuShiekah99} we conclude that between
250~K  and 140~K the holes experience a redistribution that
changes the electrical field gradients at the Ni site - a process
consistent with the formation of charge density waves or (short)
stripes.

 .

\begin{acknowledgments}
We thank Yakov Mukovskii and his coworkers of the Moscow State
Steel and Alloys Institute for the synthesis and growing of the
enriched single crystals and Ruud Hendrikx of the Technical
University of Delft for the crystal diffraction analysis. The
powder of ${\rm K_2NiF_4}$ was kindly provided by Dany Carlier
(ICMCB-CNRS, Bordeaux). Stimulating discussions with Jan Zaanen
are highly appreciated.
\end{acknowledgments}

\appendix

\section{$K$ and $\chi$ of Ni$^{2+}$}.

{\it{Dilute paramagnets}}. For experimental reasons, hardly any
NMR exists on dilute transition metal ions in a diamagnetic host,
but hyperfine fields can still be found from ENDOR data. A
didactic treatment of this procedure, using Ni$^{2+}$ as example,
has been given by Geschwind.\cite{Geschwind67} The ground term of
the free $3d^8$ ion has $L=3$, $S=1$. The sevenfold orbital
degeneracy of the free ion is lifted by the cubic crystal field
into a low-lying orbital singlet and two excited orbital triplets.
Perturbation theory of the magnetic resonance properties of
Ni$^{2+}$ considers only these three sets of states. In first
order, when only the orbital singlet is considered, the orbital
moment is quenched: the electronic $g = 2$, there is only a Curie
susceptibility $\propto 1/T$, and the NMR shift is determined by
this susceptibility and the core-polarization hyperfine field. In
second order new contributions appear: the electronic $g$-shift
$\Delta g$ (a cross term between the spin-orbit hamiltonian and
the orbital part of the electron Zeeman hamiltonian), the orbital
hyperfine field (a cross term between the spin-orbit and the
orbital hyperfine hamiltonians), the paramagnetic shielding of the
nuclear Zeeman coupling (a cross term between the orbital
hyperfine hamiltonian and the orbital part of the electron Zeeman
hamiltonian) and the $T$-independent contribution to $\chi$ (a
cross term of the orbital part of the Zeeman hamiltonian with
itself). Apart from a multiplicative constant, the paramagnetic
shielding can also be written as a product of the $T$-independent
susceptibility and the orbital hyperfine field. At this level of
perturbation, there is no dipolar part in the hyperfine
hamiltonian (but it appears in lower symmetry).

The hamiltonian for the effective electron spin $S$,
$\mathcal{H}_S$, can be written as
\begin{equation}
\mathcal{H}_S/\hbar = \beta \vec{B} \cdot (2 + \tensor{\Delta g})
\cdot \vec{S} + \vec{S} \cdot \tensor{A} \cdot \vec{I} - \gamma_I
(1 + \delta) \vec{B} \cdot \vec{I}, \label{Endorequation}
\end{equation}
where we have omitted an additional term for the zero-field
splitting, which is unimportant for our purpose,\cite{Locher63}
and the nuclear quadrupole coupling. The effective hyperfine
tensor $\tensor{A}$ now contains both the (negative) core
polarization and (positive) orbital hyperfine fields and is to a
good approximation an ``ionic" property, independent of the host.
The paramagnetic shielding is represented by $\delta$. The Knight
shift hamiltonian that corresponds to Eq.~\ref{Endorequation} is
\begin{equation}
\mathcal{H}_I /\hbar = \langle \vec{S} \rangle \cdot \tensor{A} \cdot
\vec{I} - \gamma_I (1 + \delta) \vec{B} \cdot \vec{I},
\label{Knightequation}
\end{equation}
of which Eq.~\ref{e3} is a simplified form. In that equation the
value of $\Delta g$ shows up explicitly in the expression for
$\chi_S$.

The low-spin $^{61}$Ni$^{3+}$ has been seen in MgO by ESR,
together with $^{61}$Ni$^{2+}$.\cite{Orton63,Hochli65} The
paramagnetic shielding has not be determined. For Ni$^{3+}$ the
$g$-shift is 0.17 and the absolute value of the hyperfine field
6.71 T; for Ni$^{2+}$ the values are 0.22 and 6.45 T. In general
the crystal field analysis for the $3d^7$ configuration is rather
complicated (the typical case is Co$^{2+}$, S=3/2),\cite{Fry62}
but for the low-spin configuration strong ligand field theory
applies. There is just a single electron in the $d\gamma$ shell,
which makes the magnetic resonance behavior very similar to that
of $3d^9$ Cu$^{2+}$ with a single hole in the $d \gamma$
shell.\cite{AbragamBleaney70} Even though the paramagnetic
shielding of $^{61}$Ni$^{3+}$ is not known, one may assume that in
a hypothetical NMR experiment on these dilute impurities in MgO
two clearly distinct signals would be observed for the two Ni
valencies.

{\it{Dense paramagnets}}. A typical extension of crystal field
theory to the NMR of the ``magnetic" nuclei in dense paramagnets
is provided by the discussion of $^{63}$Cu$^{2+}$ in the
YBa$_2$Cu$_3$O$_x$ system.\cite{Shimizu93} The main new
contribution that appears is the supertransferred hyperfine field,
due to the presence of the neighboring magnetic ions.
Schematically it can be thought of as due to a very small $4s$
admixture created by the super exchange; $s$ electrons have a
large positive direct hyperfine field, so even a small admixture
can cause measurable effects.

In dense paramagnetic systems like KCoF$_3$ and
CoO,\cite{Shulman59} it is found experimentally (see
Fig.\ref{figCJ1}) that the paramagnetic shielding is much smaller
than in dilute paramagnets. This is likely related to an exchange
narrowing of both spin and orbital interactions,\cite{Shulman59}
but in Eq.~\ref{Knightequation} the $\langle \vec{S} \rangle$-part
remains proportional to $\chi_S$ and $\delta$ remains
$T$-independent in the relevant $T$ range. Further support for
this hypothesis comes from the experimental data in the low-$T$
antiferromagnetic phase,\cite{Jaccarino59,Moriya59} where the
exchange becomes static and essentially the same $\delta$ as in
the dilute systems is found.

For the Ni$^{2+}$ ion the $g$-value remains close to 2 even with
an additional trigonal distortion, as in the case of an
Al$_2$O$_3$ host, where ENDOR measurements have been made.
\cite{Locher63} These data yield a ``pseudo nuclear Zeeman'' shift
of 0.028, which is comparable to the Van Vleck-shift of $\approx
0.011$ that we have measured in $S=0$ K$_2$NiF$_6$. The
$T$-independent shift in K$_2$NiF$_4$ (the extrapolations in
Fig.\ref{figCJ2}) can be expected to have a comparably small
value; all the more so because this is a dense paramagnet and
exchange narrowing (see above) might be operative.\cite{Shulman59}

We made the choice of $g=2.28$ and $S=1$, taken from data on
KMgF$_3$, to plot the dashed line in Fig. 4b. The choice of
$g$-value is somewhat arbitrary - we preferred a fluoride host for
the comparison. The values in Al$_2$O$_3$, MgO and CaO range from
2.18 to 2.33; any of these would not change our main conclusion
that a reasonable estimate for the hyperfine field would come out
at a much smaller value than in the dilute systems.The spin value
should be $S=1$, both in oxides and in fluorides.\\
The difference between the negative slope in Fig. 3b and the
positive slope in Fig. 3a comes from the $\Delta g$ effect: dilute
Mn$^{2+}$ has very nearly $\Delta g=0$, see AB70, and
therefore an almost pure core-polarization hyperfine field.

{\it{The difference between susceptibility and shift}}. When the
effects of correlations are included in $\chi$, see Ref.
\onlinecite{DeJonghMiedema01} for the original references, $\chi$
has to be written as
\begin{equation}
\chi = \frac{C}{T} \sum_{i=0}^\infty \frac{3\langle
S_{z,0}S_{z,i}\rangle}{S(S+1)}, \label{chicorrs}
\end{equation}
where $i$ runs over all spins. If there are only on-site (or
auto-) correlations (a simple paramagnet)
$\left<S_{z,0}S_{z,i}\right> = \tfrac{1}{3} S(S+1)$ and one
retrieves the Curie law.

To understand that the correlations also have an effect on the
transferred hyperfine fields it is instructive to compare the
reasoning usually applied to the cuprates with that for an ordered
antiferromagnetic phase.\cite{Mila89, Huang67, Geschwind67} In the
cuprates, correlations are neglected and therefore the transferred
field has the opposite sign to the on-site core polarization
field. In the fully correlated phase the transferred field has the
same sign. Therefore, when correlations start to grow in the
paramagnetic phase, the total hyperfine field becomes more
negative, similar to what is seen in Fig.~\ref{figCJ2}a.

\end{document}